# Hybrid III-V Silicon Photonic Crystal Cavity Emitting at Telecom Wavelengths

*Svenja Mauthe, Preksha Tiwari, Markus Scherrer, Daniele Caimi, Marilyne Sousa, Heinz Schmid, Kirsten E. Moselund, and Noelia Vico Triviño\**

IBM Research Europe, Säumerstr. 4, 8803 Rüschlikon

**Abstract**

**Photonic crystal (PhC) cavities are promising candidates for Si photonics integrated circuits due to their ultrahigh quality ($Q$)-factors and small mode volumes. Here, we demonstrate a novel concept of a one-dimensional hybrid III-V/Si PhC cavity which exploits a combination of standard silicon-on-insulator technology and active III-V materials. Using template-assisted selective epitaxy, the central part of a Si PhC lattice is locally replaced with III-V gain material. The III-V material is placed to overlap with the maximum of the cavity mode field profile, while keeping the major part of the PhC in Si. The selective epitaxy process enables growth parallel to the substrate and hence, in-plane integration with Si, and in-situ in-plane homo- and heterojunctions. The fabricated hybrid III-V/Si PhCs show emission over the entire telecommunication band from 1.2 μm to 1.6 μm at room temperature validating the device concept and its potential towards fully integrated light sources on silicon.**

The field of photonics is very promising enabling high-speed and low power optical interconnects as well as added functionalities. Especially Si photonic integrated circuits (PICs) are of great interest, allowing the use of existing low-cost and mature Si technology and enabling an ideal combination of electronics for computation and optics for high-speed data transmission [1, 2]. While efficient on-chip passive and active photonic devices have been demonstrated using Si, its indirect band gap prevents efficient light emission. To create a full optical PIC, other materials have to be integrated that enable efficient light sources like lasers and light emitting diodes (LEDs). Driven by this demand, extensive research has been carried out to integrate III-Vs on Si towards integrated light sources [3–7]. III-V materials offer a direct and tunable band gap in the telecommunication band, and hence, are suitable for the use with Si passives. The challenge however is the monolithic integration on Si due to a mismatch of the crystal lattice, polarity, and thermal expansion coefficient.

Photonic crystal (PhC) cavities have emerged as ideal candidates for small footprint integrated light sources [8, 9]. They offer ultrahigh quality ($Q$) factors combined with small mode volumes ($V$), resulting in an enhancement of the light-matter interaction at the nanoscale. Extremely high $Q/V$ ratios are featured in the so-called one-dimensional (1D) nanobeam cavities [10, 11]. Typical $Q$s on the order of $10^6$ and mode volumes approaching the diffraction limit $(\lambda/2n)^3$ are widely reported [3, 11, 12] even with low refractive index materials such as SiN [13]. This is a key advancement since it overcomes the need of freestanding structures (PhC slabs) and enables Si embedded in $SiO_2$ structures. In contrast with PhC slabs, structures with Si embedded in $SiO_2$ are mechanically more robust, can be coupled in-plane with silicon-on-insulator (SOI) waveguides and passives, and offer a more efficient thermal management [14]. PhCs also offer great design flexibility that can be exploited in many applications. They have demonstrated unique properties for the realization of lasers with high-speed modulation rates [15–17] and low thresholds [3, 18–20]. Indeed, for ideal PhC cavities thresholdless lasing has been theoretically predicted [21, 22]. The performance of PhC lasers is typically limited by poor carrier confinement and inefficient heat dissipation. The use of a small active gain medium located in the PhC cavity has proven to be very effective to address such challenges [17, 23, 24]. Different approaches have been proposed such as selective regrowth techniques to create an ultracompact buried InGaAsP heterostructure at the center of an InP PhC [17] and pick-and-place of an InAsP/InP nanowire on a Si PhC [23, 24]. While impressively demonstrating the potential of local embedding of a gain material within a PhC, fabrication issues, like the precise placement of such gain material, remain very challenging.

This work demonstrates a novel concept of a 1D III-V / Si hybrid photonic crystal cavity where the gain emitting medium (III-V) is embedded locally within a silicon photonic crystal on standard SOI wafers. The key idea of the proposed concept is to selectively replace parts of the Si PhC with active III-V materials using template-assisted selective epitaxy (TASE) [25–29]. In TASE, the III-V material is epitaxially grown from a small Si seed into a prefabricated, hollow $SiO_2$ cavity. The selective replacement of Si structures by III-Vs naturally results in a self-aligned process with high dimensional accuracy and enables direct in-plane coupling to Si waveguides. Such a growth approach enables the integration of different III-V materials [30] as well as the creation of as-grown $p$-$i$-$n$ structures parallel to the substrate [29], making ex-situ diffusion or implantation doping redundant.

Fig. 1 shows the concept of the 1D hybrid III-V / Si PhC cavity. The cavity consists of a periodic array of high refractive index rods (Si or III-Vs) embedded in a low refractive index dielectric ($SiO_2$) as depicted in Fig. 1 (b). In this work, we make use of the Si-based PhC design reported in [12] due to its favorable geometry to study and prove our hybrid concept. In principle, other designs can also be implemented with the same approach. This Si 1D PhC cavity features an ultra-high $Q$-factor ($> 10^6$) thanks to a quadratically modulated length tapering of the individual rods and can be straightforwardly coupled to waveguides. In the hybrid structure the gain material is located to overlap with the maximum of the electric field profile of the cavity mode (Fig. 1 (c)). Simultaneously, cavity losses are reduced, since the Si mirrors do not absorb in the same wavelength range as the emission. To facilitate a match to the original all-Si design [12], we extend the layout by varying the width of the III-V rods to compensate for slight differences in the refractive indices $n_{Si}$ and $n_{III-V}$ of Si and III-Vs, respectively. Thus, by adjusting the width of the III-V rods, the hybrid PhC can exhibit a similar $Q$-factor and resonance wavelength as compared to the original all-Si design. The total number of rods (Si and III-V) is kept constant (N=41) while the number of III-V rods is either 5 or 9 to overlap with the maximum or the entire mode profile, respectively (see Fig. 1(c)). Moreover, a sweep of the PhC lattice constant $a$ (distance between two adjacent rods) from 325 nm to 400 nm is included. The III-V rods consist of a

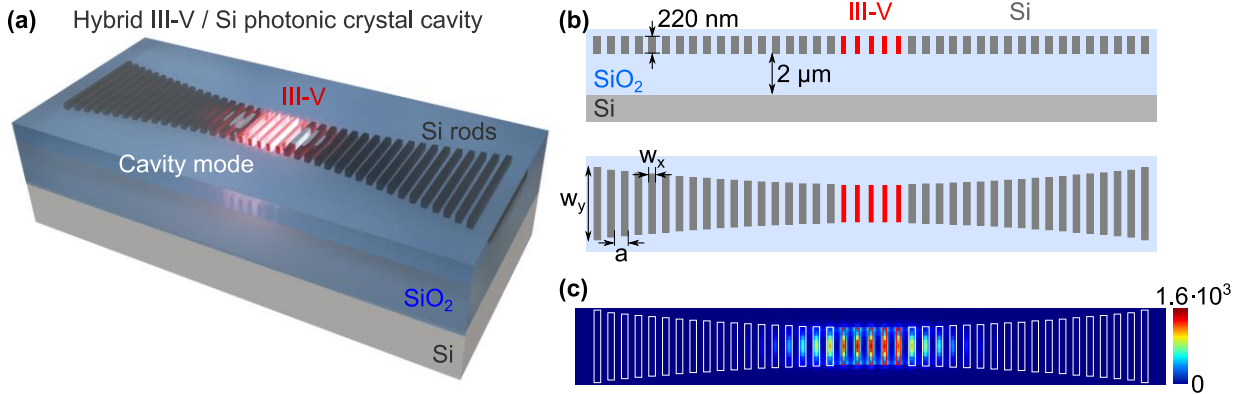

Fig. 1 (a) 3D view of the hybrid III-V / Si PhC cavity integrated on SOI. (b) Cross-section (top) and top view (bottom) schematic of the hybrid cavity. The rods are length modulated according to $w_y(i) = w_y(0) + i^2 (w_y(i_{max}) - w_y(0)) / i_{max}^2$ where $w_y(i)$ is the length of the $i^{th}$ rod. The width $w_x$ is kept constant. Both, $w_y$ and $w_x$ are proportional to the lattice constant $a$. (c) $|E_y|^2$ profile of the first cavity mode calculated via 3D FDTD.

heterostructure including InP / InGaAs / InP segments which are additionally in-situ doped during growth to form a *p-i-n* diode to resemble as close as possible an electrically actuated device configuration. This allows tuning of the emission wavelengths ranging from 1.2 µm to 1.6 µm, i.e., covering the entire telecom window.

Fig. 2 depicts the individual processing steps to create the hybrid III-V / Si PhC cavity. The fabrication starts with a commercial SOI wafer (SOITEC) with a 2 µm thick buried oxide (BOX) for sufficient optical isolation and a 220 nm thick top Si(100) layer. In a first step, the top Si layer is patterned by means of e-beam lithography and HBr dry etching. Next, SiO$_2$ is deposited, 100 nm using atomic layer deposition (ALD) and 200 nm by plasma-enhanced chemical vapor deposition (PECVD). The deposited oxide layer is planarized using chemical mechanical polishing (CMP) and reduced to a thickness of 150 nm. Next, windows are etched into the oxide layer exposing one end of the targeted Si rods in the middle of the cavity. The exposed Si is partly etched using tetramethylammonium hydroxide (TMAH). The etch rate is precise on a nm-scale and is adjusted via the temperature and concentration of the TMAH solution. Step 4 in Fig. 2 schematically depicts the partially etched Si rods and the remaining empty SiO$_2$ template. The etch is timed to match the designed length of the middle rods in the PhC cavity. The remaining Si acts as a nucleation seed during III-V growth in step 5. We note that the presence of the back-etched Si rods, which serve as growth seeds, and the oxide window to reach the rods slightly break the design symmetry of the PhC. This, however, is not a limiting factor of the concept and can be optimized by e.g. performing a second lithography. Metal-organic chemical vapor deposition (MOCVD) is used to grow III-V materials selectively from the Si seeds into the empty SiO$_2$ templates (Fig 2, step 5). Shortly before the growth, a 20 s dilute hydrofluoric acid (DHF) etch is performed to remove native oxide on the Si seeds, during which the SiO$_2$ cavity is enlarged by 30 nm. The *n*-InP / *i*-InGaAs / *p*-InP / *p*-InGaAs heterojunction are grown at 550°C using TMIn, TBP, TBAs, TMGa with DEZn and TESn as *p*- and *n*-type dopants. The additional outer *p*-InGaAs segment was added to mimic a future contact layer to the *p*-InP segment. The fabrication of electrical contacts is out of scope of this initial demonstration and might require an optimization of the hybrid PhC cavity.

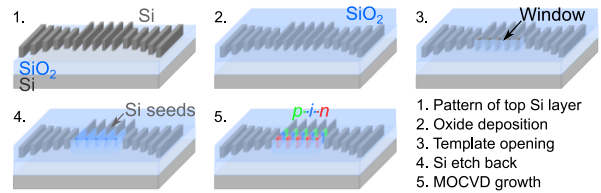

Fig. 2 Fabrication steps during TASE process. Starting from a standard SOI wafer, the top Si is patterned according to the PhC design (1.). 2. Oxide is deposited, and the sample is planarized. 3. Template opening exposing the underlying Si for the center rods. 4. Partial etch of the Si rods in the center. 5. MOCVD growth of n-i-p InP / InGaAs heterostructure.

Fig. 3 (a) and (b) depict scanning electron microscopy (SEM) overview images of two hybrid PhC cavities with 5 and 9 III-V rods, respectively. A substantial variation in the length of the grown III-V rods is observed, which is an indication of non-uniform nucleation and likely due to residues on the Si surface. This is not intrinsic to the growth method. Fig. 3 (c) shows an SEM top-view image of a single III-V rod. Fig. 3 (d) depicts a cross-section along the center of the PhC cavity which demonstrates the hybrid integration and illustrates the alignment of the III-V rods with the Si rods. To further investigate the material quality of the grown devices, scanning transmission electron microscopy (STEM) analysis is performed as shown in Fig. 3 (e) and (f). The high-resolution images confirm a single crystalline material and epitaxy relationship with the Si layer.

The hybrid PhC cavities are characterized by micro-photoluminescence (µ-PL) spectroscopy. The cavities are illuminated with an 850 nm ps-pulsed supercontinuum laser (repetition rate: 78 MHz, spot size: ~1 µm) to optically excite the grown III-V material. Both, the excitation and response of the sample, are focused and collected from the top using a ×100 objective (NA 0.6). The optical response is characterized using a spectrometer with a liquid nitrogen cooled InGaAs linear array photodetector. The PhC structure exhibits a spontaneous emission background starting from 1.2 µm to 1.6 µm along with the different cavity modes. The gain emitting material allowed us to observe cavity modes over the entire telecom window, by changing the lattice constant *a* of the PhC. Fig. 4 (a) shows the PL spectra of hybrid PhC cavities with 5 III-V rods and varying *a*. Three cavity modes are visible. Fig. 4 (b) depicts the measured center wavelength versus *a* of the optical modes for hybrid PhC cavities with 5 and 9 rods in the center of the hybrid cavity. The resonances shift to longer wavelengths with increasing *a* demonstrating full tunability over the telecom range and hence, enabling emission in the relevant O- and C-bands. It is worth mentioning that despite the non-uniform length of the III-V rods, we observe similar behavior for all the measured cavities, with similar experimental *Q*-factors on the order of $10^3$. This is illustrated in Fig. 4 (b) for 17 different structures. No significant differences were observed with regard to the number of III-V rods (5 or 9). We believe that the observed robustness and high yield in such non-perfect PhC cavities relies on

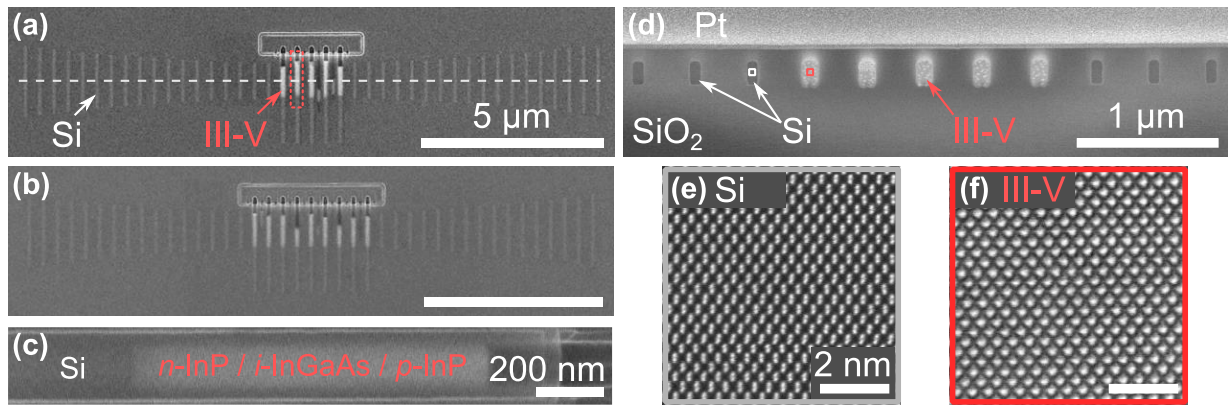

*Fig. 3 (a) SEM images of hybrid PhC cavity with 5 (a) and 9 (b) III-V rods surrounded by Si mirrors. (c) SEM image of a single III-V rod marked in (a). (d) STEM cross-section of a hybrid PhC cavity with 5 rods along the x-axis of the hybrid PhC cavity (dashed line in (a)). The III-V rods are aligned in between the Si rods of the hybrid PhC cavity. (e) & (f) High resolution STEM images of the Si (e) and grown III-V (f) rods at marked positions in (d). Although only the first III-V rod is selected here, it is representative of the 4 remaining ones.*

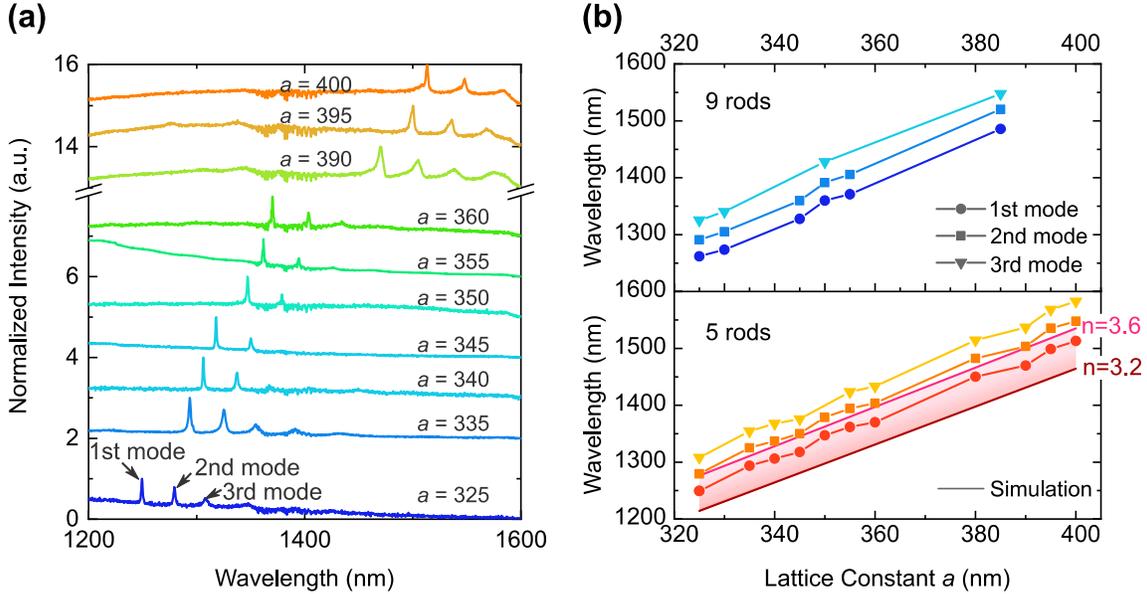

Fig. 4 (a) PL spectrum of hybrid PhC cavities with 5 rods for increasing lattice constant a ($w_x = 0.33a$, $w_y = 5a$). (b) Measured peak wavelength of the three resonant modes versus lattice constant of the hybrid PhC plotted for two different number of III-V rods: 5 and 9 in the hybrid cavity ($w_x = 0.33a$, $w_y = 5a$). One can observe cavity resonances spanning the entire telecom spectral range. The shadow area encloses the peak wavelengths of the first cavity mode obtained by 3D-FDTD simulations using refractive index values of $n_{III-V}$ between 3.2 and 3.6.

the local integration of III-Vs within the central part of an otherwise ideal Si lattice of our hybrid PhC.

We perform 3D FDTD simulations using *Lumerical FDTD Solutions*. Fig. 4(b) depicts the simulated wavelength shift over lattice constant of the first cavity mode. The lower and upper limits of the shadow area correspond to the resonance wavelengths for constant values of the III-V refractive index of 3.2 and 3.6, respectively [31]. Owing to the complexity of our III-V rods combining InP and InGaAs, estimating an exact refractive index dispersion curve or value is not trivial. The simulated trend (shadow area) agrees well with the experimental data (red circles).

Fig. 5 (a) depicts the spectra of a hybrid PhC cavity under increasing optical excitation. With increasing excitation fluence, the light intensity of the resonances as well as the spontaneous emission background, increase. Fig. 5 (b) shows the integrated optical power (light in-light out (LL) curve) of the cavity mode at 1368.5 nm for increasing pump powers. Fig. 5 (c) depicts the full width at half maximum (FWHM) of the resonant peak as well as the LL curve in log-log scale. The FWHM decreases from ~5.6 nm at low excitation powers to ~1.8 nm. The slight increment in the FWHM at very high excitation fluence is observed due to plasma dispersion effect which result in a blueshift of the peak [32, 33].

We believe that the observed linewidth narrowing and measured LL curve show evidence of lasing in our structures at room temperature. However, to unambiguously demonstrate lasing in such structures, further studies of carrier lifetime and photon statistics (correlation measurements) are required. This goes beyond the scope of this proof-of-concept paper.

In summary, we demonstrated a novel method to create a hybrid III-V / Si light source on SOI with tunable operation over the entire telecom band at room temperature. The hybrid PhC crystal concept along with the presented fabrication technique benefits from the use of mature Si processing technology and existing Si PhC designs. The local

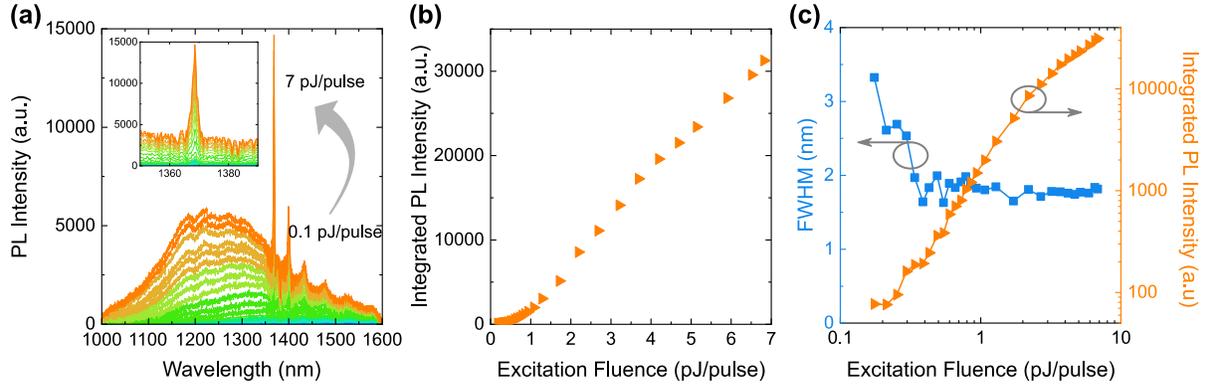

Fig. 5 (a) Measured PL intensity of a hybrid PhC for increasing excitation fluences ($a = 375$ nm, 5 rods, $w_x = 0.3a$, $w_y = 5a$). (b) Integrated PL intensity versus excitation fluence of the resonant peak at 1368 nm in (a). (c) FWHM and integrated PL intensity from (b) versus excitation fluence in double log scale.

embedding of the III-Vs on Si takes place seamlessly in the very last step of the fabrication process. Thus, one can foresee fabrication of the full Si PhC in a silicon foundry together with Si waveguides and passives. The III-V gain material was implemented as lateral heterostructures with an in-situ as-grown *p-i-n* doping profile [34] to illustrate a path towards electrically driven, integrated light source on silicon.

**Corresponding Author**

* nvi@zurich.ibm.com

**Author Contributions**

S.M., P.T., M.S., H.S., K.M., and N.V.T. created the device concept. N.V.T. performed FDTD simulations. S.M. and P.T. fabricated the sample with support of D.C. for the CMP. S.M. performed the optical characterization on the devices. S.M., N.V.T., and P.T. analyzed the data. M.Sousa performed the FIB lamella and STEM characterization. K.M., and N.V.T. lead and managed the project. All authors discussed the results. The manuscript was written by S.M. and N.V.T., with contributions of all authors, and all authors have given approval to the final version of the manuscript.

**Funding Sources**

This work received financial support from H2020 ERC project PLASMIC, Grant #678567 and Swiss National Science Foundation Spark project SPILA, Grant #CRSK-2_190806.

ACKNOWLEDGMENT

The authors gratefully acknowledge the BRNC staff for technical support.

ABBREVIATIONS

PhC photonic crystal, TASE template-assisted selective epitaxy, PIC photonic integrated circuits, LED light emitting diode, SOI silicon-on insulator, BOX buried oxide, ALD atomic layer deposition, PECVD plasma-enhanced chemical vapor deposition, CMP chemical mechanical polishing, TMAH tetramethylammonium hydroxide, MOCVD metal-organic chemical vapor deposition, DHF dilute hydrofluoric acid, TMIn Trimethylindium, TBP Tertiarybutylphosphine, TBAs Tertiarybutylarsine, TMGa Trimethylgallium, DEZn Diethylzinc, TESn Tetraethyltin, SEM scanning electron microscopy, EDX energy-dispersive X-ray spectroscopy, STEM scanning transmission electron microscopy, FFT Fast Fourier transform, PL photoluminescence, NA numerical aperture, LL light in-light out, FWHM full width at half maximum, FDTD finite-difference-time-domain.